\input harvmac

\vskip 1cm

 \Title{ \vbox{\baselineskip12pt\hbox{  Brown Het-1251, CTP-MIT-3069}}}
 {\vbox{
\centerline{  On the polarization of unstable D0-branes    }
\centerline{  into non-commutative odd spheres  }  }}

\centerline{$\quad$ { Zachary Guralnik } }
\smallskip
\centerline{{\sl  Center for Theoretical Physics,  }}
\centerline{{\sl Massachusetts Institute of Technology }}
 \centerline{{\sl Cambridge, Massachusetts 02139 }}
\medskip
\centerline{$\quad$  { Sanjaye Ramgoolam }}
\smallskip
\centerline{{\sl Department of Physics}}
\centerline{{\sl Brown  University}}
\centerline{{\sl Providence, RI 02912 }}
\smallskip
\centerline{{\tt zack@mitlns.mit.edu, ramgosk@het.brown.edu}}
 \vskip .3in

 We consider the polarization of unstable type IIB D0-branes
 in the presence of a background  five-form field
 strength. 
 This phenomenon is studied from the point of view 
 of the leading terms in the non-abelian 
 Born Infeld action of the unstable D0-branes. 
 The equations have $SO(4)$ invariant solutions 
 describing a non-commutative 
 $3$-sphere, which becomes 
 a classical $3$-sphere in the large $N$ limit. 
 We discuss the interpretation of these solutions 
 as spherical D3-branes. 
 The tachyon plays a tantalizingly geometrical role in relating
 the fuzzy $S^3$ geometry to that of a fuzzy $S^4$.

%\draftmode 
 
\lref\bfss{T. Banks, W. Fischler , S. Shenker , L. Susskind ``M Theory
as a Matrix Model: a Conjecture,''
Phys.Rev.{\bf D55}, (1997) 5112, hep-th/9610043.} 
\lref\grt{ O. Ganor, S. Ramgoolam, W. Taylor, IV, ``Branes Fluxes and Duality
in M(atrix) Theory,'' Nucl.Phys.{\bf B492} (1997) 191, hep-th/9611202.}
\lref\bss{ T. Banks, N. Seiberg, S. Shenker, ``Branes from Matrices,''
  Nucl.Phys.{\bf B490} (1997) 91-106, hep-th/9612157.}
\lref\myers{R. Myers, ``Dielectric Branes,'' JHEP {\bf 9912} (1999) 022, 
hep-th/9910053  } 
\lref\antram{ A. Jevicki and S. Ramgoolam, `` Noncommutative Gravity from the 
ADS/CFT correspondence,'' JHEP {\bf 9904} (1999) 032, hep-th/9902059.}   
\lref\hrt{  P.M.Ho, S.Ramgoolam and R.Tatar, ``Quantum Space-times 
and Finite N
Effects in 4-D Superyang-mills Theories,'' Nucl.Phys.{\bf B573} (2000) 364,
hep-th/9907145.}
\lref\stein{  Steinacker .. ?   }
\lref\cola{  A. Connes, G. Landi
``  Noncommutative Manifolds the Instanton Algebra 
and Isospectral Deformations,'' math.QA/0011194 }
\lref\kluson{ J. Kluson, ``Branes from N Non-BPS D0 Branes,'' JHEP {\bf 0011}
 (2000) 016, hep-th/0009189.} 
\lref\fadd{ faddeev .. ?  } 
\lref\cas{ J.Castelino, S. Lee and W. Taylor IV, ``Longitudinal Five-Branes
as Four Spheres in Matrix Theory,'' Nucl.Phys.{\bf B526} (1998) 334, 
hep-th/9712105.}
\lref\holi{ P.M.Ho and M.Li, ``Fuzzy Spheres in ADS/CFT Correspondence
and Holography from Noncommutativity,'' hep-th/0004072.  } 
\lref\hor{ Horava ; perhaps others }  
\lref\malstrom{ J. Maldacena and A. Strominger, ``ADS(3) Black Holes and
a String Exclusion Principle,'' HEP {\bf 9812} (1998) 005, hep-th/9804085.}
\lref\hidclass{ A. Jevicki, M. Mihailescu and  S. Ramgoolam,
``Hidden Classical Symmetry in Quantum Spaces at Roots of Unity: from Q
Sphere to Fuzzy Sphere, ''
hep-th/0008186. } 
\lref\hamo{ J. Harvey and G. Moore, ``Noncommutative Tachyons and
K-Theory,'' hep-th/0009030.}
\lref\fro{ J. Frohlich, O. Grandjean and A. Recknage, 
``Supersymmetric Quantum Theory and (Non-commutative) 
Differential Geometry,'' Commun.Math.Phys.{\bf 193} (1998) 527, 
hep-th/9612205    } 
\lref\frt{ L. Faddeev, N. Reshetikhin and L.Takhtajan, ``Quantization of Lie
Groups and Lie Algebras,'' Lengingrad Math.J.{\bf 1} (1990), 
Alg.Anal.{\bf 1} (1989) 178.}
\lref\ho{ P. M. Ho , ``Fuzzy Sphere from Matrix Model,'' 
JHEP {\bf 0012} (2000) 014,
hep-th/0010165. } 
\lref\bartmin{ J. Minahan and B. Zwiebach, ``Effective Tachyon Dynamics in
Superstring Theory,'' hep-th/0009246.}
\lref\marmo{ D. Kutasov, M. Marino and G.Moore, ``
Remarks on Tachyon Condensation in Superstring Field Theory.''
hep-th/0010108.}
\lref\muk{ S. Mukhi and N. Suryanarayana, ``Chern-Simons Terms on
Noncommutative Branes,'' JHEP {\bf 0011} (2000) 006, hep-th/0009101.} 
\lref\rad{ R. Tatar, `` T-duality and Actions for Non-Commutative D-Branes,''
 hep-th/0011057.} 
\lref\wil{ C. Kennedy and A. Wilkins, ``Ramond-Ramond Couplings
on Brane-Anti-Brane Systems,'' Phys.Lett.{\bf B464} (1999) 206,
hep-th/9905195. }
\lref\janmees{ B. Janssen and P. Meessen, ``A Nonabelian Chern-Simons
Term for NonBPS D-Branes,''  hep-th/0009025.  } 
\lref\berg{ 
E.A. Bergshoeff, M. de Roo, T.C. de Wit, E. Eyras, and S. Panda,
``T Duality and Actions for NonBPS D-Branes,'' 
JHEP {\bf 0005} (2000) 009,
hep-th/0003221. } 
\lref\horav{ P. Horava, 
``Type IIA D-Branes, K-Theory, and Matrix Theory,'' 
Adv.Theor.Math. Phys.{\bf 2} (1999) 1373, hep-th/9812135.}
\lref\bcr{ M. Billo, B. Craps and F. Roose,  
``Ramond-Ramond couplings of non-BPS D-branes,'' JHEP {\bf 9906}
(1999) 033,  hep-th/9905157. } 
\lref\hhsphal{ J. Harvey, P. Horava and P. Kraus,
 ``D Sphalerons and 
the Topology of String Configuration Space,'' JHEP {\bf 0003}
(2000) 021, hep-th/0001143.}
\lref\dgsphal{ N. Drukker, D. Gross and N. Itzhaki, 
``Sphalerons, Merons 
and Unstable Branes in ADS,'' Phys.Rev.{\bf D62} (2000)
086007, hep-th/0004131.}
\lref\mad{J. Madore, ``The Fuzzy Sphere,'' Class.Quant.Grav.{\bf 9}
(1992) 69.}
\lref\sen{A. Sen, ``Descent Relations among Bosonic D-Branes,''
Int.J.Mod.Phys.{\bf A14} (1999) 4061, 
``Stable non-BPS bound states of BPS D-branes,''
JHEP {\bf 9808} (1998) 010, hep-th/9805019; ``SO(32) Spinors of Type I
and other Solitons on Brane-antibrane Pair,'' JHEP {\bf 9809} (1998) 023,
hep-th/9808141; ``Type I D-particle and its Interactions,'' JHEP {\bf 9810}
(1998) 021, hep-th/9809111;
``NonBPS States and Branes in String Theory,''
hep-th/9904207, and references therein.}
\lref\garousi{ M. Garousi, ``Tachyon Coupling on non-BPS D-branes and 
Dirac-Born-Infeld action,''Nucl.Phys.{\bf B584} (2000) 284-299, 
hep-th/0003122. }
\lref\sena{ A. Sen, ``Supersymmetric World-volume Action for Non-BPS
D-branes,''JHEP {\bf 9910} (1999) 008, hep-th/9909062.}
\lref\klusonaction{ J. Kluson, ``Proposal for Non-BPS D-Brane Action,''
Phys.Rev. {\bf D62} (2000) 126003, hep-th/0004106.}
\lref\witk{E. Witten, ``D-Branes and K Theory,'' JHEP {\bf 9812} 
(1998) 019, hep-th/9810188.}
\lref\bartmintwo{ J. Minahan and B. Zwiebach, ``Gauge Fields and Fermions
 in Tachyon Effective Field Theories,'' hep-th/0011226.}
\lref\bartminthree{ J. Minahan and B. Zwiebach, 
``Field Theory Models of Tachyon and Gauge Field String Dynamics,''
JHEP {\bf 0009} (2000) 029, hep-th/0008231.}
\lref\mst{J. McGreevy, L. Susskind and N. Toumbas, 
``Invasion of the Giant Gravitons from Anti-de Sitter Space,''
JHEP {\bf 0006} (2000) 008, hep-th/0003075.}

\Date{ 12/2000 }

\newsec{ Introduction }

 The view that zero-branes are fundamental
 objects that capture a number of important 
 features of M-theory leads one to expect 
 that higher branes can be constructed from 
 zero branes in Type IIA string theory \refs{\bfss,\grt,\bss}.
  Zero branes of Type IIA can be polarized 
  into D2-branes in the presence of background 
  Ramond-Ramond flux \myers, leading to a natural physical context 
  where the non-commutative 2-sphere \mad\ appears.
 The equations defining the non-commutative 2-sphere appear naturally 
 as conditions obeyed by the matrices corresponding 
 to transverse coordinates  in the worldvolume 
 theory of the zero-branes.   
In this paper,  we will instead consider the polarization of
 unstable type IIB D0-branes by Ramond-Ramond flux.  
This provides a natural physical  context 
  for obtaining odd non-commutative spheres. 

Recently, much progress has been made in understanding
the dynamics of unstable branes. One of the remarkable
features of these systems is that  
stable branes can be obtained as solitons of 
higher dimensional unstable branes 
\refs{\sen, \witk, \horav},  leading to an interpretation of
brane charges in terms of K-theory. 
A Non-abelian Born-Infeld type action for unstable D-branes has
been written down \refs{ \sena, \garousi, \berg,  \bartmintwo, 
\bartmin, \bartminthree
\klusonaction }
as have their Chern-Simons couplings to Ramond-Ramond forms
\refs{\sena, \horav, \bcr, \wil, \janmees, \muk}. 
This action is for the most part phenomenological, 
satisfying certain physical requirements. Its exact form is 
unknown, although the tachyon potential and kinetic term have
been computed exactly within the context of boundary string 
field theory \marmo, giving a result agreeing with a 
proposal in \bartmin. 
Using the nonabelian action of unstable D0-branes, 
solutions corresponding to higher dimensional branes
with a flat geometry
were obtained in \kluson.
In the presence of a constant Ramond-Ramond 
five-form field strength,  the D0 action gives  
a system of matrix equations which
one might expect to have unstable solutions 
describing a fuzzy three-sphere. 

We will find matrices $\phi^i$, for $i = 1 \cdots 4$, 
which define a fuzzy $3$-sphere and 
solve the D0 equations of motion in the five-form background. 
These matrices satisfy $\phi^i\phi^i = R^2$ and commute in
the large $N$ limit. 
Our construction is based on that which gave
the fuzzy $4$-sphere of \cas. The fuzzy $S^3$ may be interpreted
as a subspace of the fuzzy $S^4$.
The solution for the tachyon is is related to the fifth 
embedding coordinate involved in the definition of the fuzzy
$S^4$.  These solutions exist for matrices of size
$N = { ( n+ 1 ) ( n+3 ) \over 2 }$, 
where $ n$ is an odd number.

At finite $N$,  the coupling of our solutions to the five-form 
field strength resembles the dipole moment coupling of a 
spherical D3-brane. However, in the large $N$ limit 
where the non-commutative 
$S^3$ becomes a classical $S^3$,  this dipole moment vanishes
for the particular class of large $N$ solutions we consider.
Thus we do not interpret these solutions
as spherical D3-branes. We expect that this 
situation may change if one considers a consistent 
supergravity background, instead of a flat geometry with five-form 
flux. Furthermore, if a dual D3-brane description exists, then
the phenomenological Lagrangian which we
work with might be insufficient to obtain the correctly 
normalized couplings,  which depend sensitively on tachyon 
dynamics.

\newsec{Unstable D0-brane action}

We are interested in the behavior of an unstable IIB D0-brane
in a constant background RR five-form field strength
\eqn\ff{ 
 H^{(5)} =  h dt \wedge dx^1 \wedge dx^2 \wedge dx^3 \wedge dx^4
        - h dx^5 \wedge dx^6 \wedge dx^7 \wedge dx^8 \wedge dx^9
} 
The low energy action for an unstable D0-brane in type IIB
string theory can be obtained
by T-dualizing the unstable IIA D9-brane action as in \myers. 
There is a large literature on  Born-Infeld and Chern-Simons 
 terms for unstable branes
\refs{\sena, \garousi, \berg,  \bartmintwo, 
\bartmin, \bartminthree, \horav, \bcr, \wil, \janmees, \muk}.
The D9 action is $S^9 = S^9_{DBI} + S^9_{CS}$ where
\eqn\dn{ 
S^9_{DBI} = {1 \over g_s {\alpha^{\prime}}^5 (2\pi)^9}
        \int d^{10}x Tr \left[ V(T)
        \sqrt{det( G + 2\pi\alpha^{\prime} F) } 
        + \alpha^{\prime} f(T)\sqrt{det G}D_{\mu}T D_{\mu}T \right]
} 
Following conjectures of Sen \sen,  one is led to expect
that  $f(T)$ and $V(T)$ vanish at the global minimum of $V(T)$, 
for a suitable choice of the variable $T$.
If one changes variables such that the tachyon kinetic term is 
canonically normalized, or $f(T) = 1$,  then the 
Chern-Simons coupling is of the form : 
\eqn\cst{ 
S^9_{CS} = \int Tr DT \wedge C \wedge e^{F+B}
}
A form for the potential and kinetic terms was proposed
in \bartmin\ and computed exactly within the context of 
boundary string field theory in \marmo,  giving  
\eqn\kinet{V(T) = \mu f(T)}
with 
\eqn\pot{V(T) = e^{-T^2/4}}
and 
\eqn\moo{\mu = {1\over 2 ln 2}.}
We shall find solutions for this particular potential and kinetic term,
however the existence of fuzzy three-sphere solutions does not
depend on the detailed form of the potential. 
With a choice of variables such that the tachyon kinetic term is 
given by \kinet\pot\ ,
the Chern-Simons coupling
is\foot{ We thank Barton Zwiebach for pointing this out.}   
\eqn\cst{ 
S^9_{CS} = \int Tr \sqrt{2ln2} e^{-T^2/8} DT \wedge C \wedge e^{F+B}
.}

Upon T-dualizing in all 9 spatial directions,
one finds the D0 action in the presence of the above five-form
field strength is $S^0 = S^0_{DBI} + S^0_{CS}$ 
where
\eqn\dbit{\eqalign{ 
S_0^{DBI} =  {1\over g_s {\alpha^{\prime}}^{1/2}} 
        \int dt STr  V(T)
        \sqrt{  det (\delta_{ab} + 2\pi\alpha^{\prime}[X^a,X^b])} 
        &\cr
        - \alpha^{\prime} f(T) [X^a,T][X^a,T]  
        + \cdots  
        }} 
and 
\eqn\thecs{ 
S_0^{CS} = {\alpha^{\prime}}^{3/2} h \int dt STr 
        e^{-T^2/8} \left[ [X^l, T] 
	\epsilon_{ijkl}X^i X^j X^k \right], 
} 
where $STr$ indicates a symmetrized trace. 
The indicies $i,j,k,l$ run from $1$ to $4$ and label the
matrix coordinates in which the fuzzy $S^3$ will be embedded. 
We have absorbed numerical factors of order $1$ in the 
definition of $h$. 
Since we will look for a static solution,  time derivative terms have been
dropped. 
% There are also terms in $S_0^{CS}$ 
%involving $A_0$ which may be set to zero
%after imposing the Gauss law constraint. 
%
%\eqn\gsl{ 
%        {\alpha^{\prime}}^{3/2} h \int dt Tr 
%        \left( [A_0,T] \epsilon_{01234mnpqr} X^m X^n X^p X^q X^r 
%        + [X^m,T] \epsilon_{01234mnpqr} [A_0,X^n] X^p X^q X^r \right)
%}
%We shall consider only solutions with $X^m =0$ for $m=5 \cdots 9$, so
%one can neglect the contribution to the Gauss law coming from this
%term.

In terms of dimensionless adjoint scalars,
  $\phi^i = \sqrt{\alpha^{\prime}} X^i$,
 and keeping only the 
leading terms in the DBI action for$[\phi^i,\phi^j] << 1$,
we have: 
\eqn\flac{\eqalign{   
  S = - { 1 \over g_s} \int dt Str V(T) 
\bigl(   1  - { 1\over 4 } [\phi^{i},
  \phi^{j} ]   [\phi^{i}, \phi^{j} ] \bigr ) - 
 \bigl(  f(T) [ \phi^{i}, T ]
  [ \phi^{i}, T ] \bigr )& \cr 
   - h e^{-T^2/8} \phi^{i} \phi^{j} \phi^{k} [ \phi^{l}, T] 
	\epsilon_{ijkl}  &\cr 
}}
We shall find a class of solutions for which 
$T^2$ is proportional to the identity, and $\{T, \phi^i\} = 0$.
In this case, the equations of motion are
\eqn\teq{\eqalign{  
V^{\prime}(T)(1- {1 \over 4}[\phi^i,\phi^j][\phi^i,\phi^j]) - 
f^{\prime}(T)[\phi^i,T][\phi^i,T] 
 & \cr 
- \left[ [\phi^i,T] f(T), \phi^i \right]
- \left[ f(T)[\phi^i,T], \phi^i \right]  &\cr 
-2h e^{-T^2/8} \epsilon_{ijkl}\phi^i\phi^j\phi^k\phi^l 
&\cr }}
and

\eqn\peq{\eqalign{ 
\left[[\phi^i, T] f(T),T \right] + 
\left[ f(T)[\phi^i,T],T \right] 
 &\cr 
+ {1 \over 2} \left[ [\phi^i,\phi^k] V(T), \phi^k \right] + 
{1\over 2} \left[ V(T) [\phi^i,\phi^k], \phi^k \right]  
&\cr 
- h e^{-T^2/8} \epsilon_{ijkl} \left( [\phi^j,\phi^k][\phi^l,T] 
+ [\phi^l,T][\phi^j,\phi^k] \right)  = 0.
}}
Note that the Chern-Simons contribution to the above equations
is modified if one seeks a more general class of solutions.

\newsec{Non-commutative $S^3$ solutions at $N=4$ }
We will initially consider $N =4$ and seek solutions with the ansatz
\eqn\ans{ \phi^i = a\gamma^i + i b\gamma^i\gamma^5.} 
and
\eqn\tch{T  = d\gamma^5, }
where the $\gamma^i$ are $4 \times 4$ hermitian 
matrices satisfying the $Spin(4)$ 
Clifford algebra, and  $a$, $b$ and $d$ are real numbers. 

Note that there is a more general $Spin(4)$ invariant class of 
solutions having
\eqn\tchgen{T = c+d\gamma^5,}  
For this  the ansatz for $T$, one has 
\eqn\fnfp{\eqalign{ 
f(T) &= q + r\gamma^5, \cr 
f^{\prime}(T) &= \omega + \lambda \gamma^5. \cr 
}}
where the quantities $q,r,\omega$ and $\lambda$ depend on $c$ and $d$ 
in a manner determined
by the form of the potential. More explicitly; 
\eqn\exprs{\eqalign{ 
q &=  1/2 ( f(c+d ) +  f(c-d) )  \cr 
r &=  1/2 ( f(c+d) - f(c-d) )  \cr 
\omega &=  1/2 ( f'(c+d ) +  f'(c-d) ) \cr 
\lambda &=  1/2 ( f'(c+d ) -  f'(c-d) ) \cr 
}}

Since we restrict ourselves here to the class of solutions
for which for which $c=0$, and since $f(T) = f(-T)$,
one has $r= 0$ and 
\eqn\vacc{\omega = tr f^{\prime}(T) = tr V^{\prime}(T) =0.}
Writing $V(T) = \mu f(T)$
and inserting the above ansatz
into the equations of motion gives: 

\eqn\wkeq{\eqalign{ 
& \lambda \gamma^5 
\left( \mu + {3 \over 4} \mu R^4 + 4 d^2 R^2 \right) \cr
+ & \left( 8 dq R^2  -   3h e^{-d^2/8} R^4 \right) 
\gamma^5 = 0. \cr}} 
for the $T$ equation of motion, and 
\eqn\wkeqi{ 
\left( d^2q + {3 \over 8}\mu q R^2 - {3 \over 2} 
h d R^2 \right)(a\gamma^i + ib\gamma^i\gamma^5) = 0.
}
for the $\phi^i$ equations of motion,
where
\eqn\radius{R^2 = \phi^i\phi^i = 4(a^2 + b^2).} 
is the squared radius of the fuzzy $S^3$.
%The coefficient of $\omega $ cannot 
%be zero for positive $R^2, \mu, d^2$. 
%This requires  $\omega = 0$,  or 
%
%This gives a relation between $c$ and $d$, which, in general, 
%  depends on the form
%of the potential. Given our ansatz for the tachyon, and equation
%\exprs\ 
% we see that for $c=0$, the above
% reduces to $$ V^{\prime} (d) - V^{\prime } (-d). $$
% This is automatically satisfied if $V$ is a symmetric
% function under $ T \rightarrow -T$  as it should be in this 
% context, see for example \horav. 
For solutions of this form note that 
 we can write the above equations of motion as : 
\eqn\intfm{\eqalign{  
\left( \mu\lambda + {3 \over 4} \lambda \mu R^4 + 4 d^2 R^2 \lambda 
+ 8 dq R^2  -  3he^{-d^2/8} R^4 \right )   { T \over d }  &= 0\cr 
\left( d^2q + {3\over 8}\mu q R^2 - {3 \over 2} 
h e^{-d^2/8} d R^2 \right)  X^{i} & =0 \cr}}
Since $c=0$, 
\eqn\czero{\eqalign{ & \mu q = e^{-d^2/4}, \cr 
 & \mu \lambda = -{d\over 2} e^{-d^2/4}.}}
Thus if one defines 
\eqn\dfh{{\hat h} = e^{d^2/8} h} 
The equations of motion become,
\eqn\newq{\eqalign{ -{d\over 2} - {3\over 8} d R^4 
-2{d^3\over \mu}R^2 + 8 {d\over \mu} R^2 - 3 {\hat h} R^4 = 0 & \cr
{d^2\over \mu} +{3\over 8} R^2 - {3\over 2} {\hat h} d R^2 = 0 &\cr}}
These two equations can be solved for $d(h)$ and $R(h)$.  The physical
solutions require real positive $R^2$. 

The energy of the solutions is given by 
\eqn\eng{ E = 4 q  ( \mu + { 3 \over 4 } \mu R^4 + 4 d^2 R^2 ) - 48 h
d R^4 }  
The solutions of interest here 
are unstable. Therefore unlike the polarization 
of $D0$ into $D2$ discussed in \myers\ solutions 
of interest may have higher energy than 
the trivial solution with $ \phi^{i} = 0 $. 
Both the energies and couplings to 
Ramond-Ramond forms  of these
solutions are subject to modifications 
by higher order terms in the action, which may 
have to be taken into account in attempts 
to compare the large $N$ solutions with D3-branes.
We will discuss solutions of the large $N$ generalization
of these equations in section 5.

\newsec{ Fuzzy $S^3$ for general $N$ } 

One can generalize the above solutions to larger matrices
in a manner which gives a commutative $S^3$ in the large $N$ limit.
This can be done using methods similiar to those used in the construction 
of the fuzzy $S^4$ \cas,  with a few crucial differences.
In particular,  it neccessary to embed the solutions 
in certain {\it reducible} representations of 
$Spin(4)$.  In some sense,  the fuzzy $S^3$ which we 
obtain can be viewed as a subspace of the fuzzy $S^4$    

For the fuzzy $S^4$,  the matrices $G_{\mu}$ satisfying 
$G_{\mu}G_{\mu} = R^2$ are embedded in the irreducible symmetric tensor  
representations of $Spin(5)$;    
%%% ( See if absence of lower brane charges requires this choice )
%%%  QUES: How does it compare with the dimension of module algebras 
%%%       coming from quantum groups ? 
\eqn\fuzfr{ G_{\mu} = \bigl( 
\Gamma_{\mu} \otimes 1 \otimes 1 \cdots \otimes 1 
                       + 1 \otimes \Gamma_{ \mu } \otimes 1 \cdots
\otimes 1 \cdots + 1 \otimes \cdots \otimes 1 \otimes \Gamma_{\mu }
\bigr)_{sym} } 
The index $\mu $ runs from $1$ to $5$. 
It is convenient to rewrite this 
as 
\eqn\fuzfr{ G_{\mu} = \sum_{k} \rho_{k} ( \Gamma_{\mu } ) P_{n} } 
 where the right hand side is a set of operators 
 acting on the n-fold tensor product $V \otimes V \cdots V$. 
 The expression  $  \rho_{k} ( \Gamma_{\mu } ) $ is the 
 action of $ \Gamma_{\mu } $ on the $k$'th factor of the tensor 
 product. 
\eqn\defr{
 \rho_{k} ( A )  | e_{i_1} e_{i_2} \cdots e_{i_k} \cdots e_{i_n} > 
   =  A^{j_k}_{i_k}  | e_{i_1} e_{i_2} \cdots e_{j_k} \cdots e_{i_n} >
} 
The symmetrization operator  $P_n$ is given by 
 $ P_n = \sum_{\sigma \in S_n} { 1 \over n!} \sigma $
where $\sigma $ acts as : 
\eqn\defp{
 \sigma   | e_{i_1} e_{i_2} \cdots e_{i_k} \cdots e_{i_n} > 
   =   | e_{i_ {\sigma(1)} } e_{i_{\sigma(2)}}  \cdots 
e_{i_{\sigma(n)}} >  }

We wish to construct a fuzzy $S^3$ with manifest $Spin(4)$ covariance. 
As a first step in constructing a fuzzy $S^3$,  we will consider
the matrices $G_i$ defined above,  with index $i$ running from $1$
to $4$.  The symmetric tensor representations of $Spin(5)$
decompose under $Spin(4)$ into a sum of reducible representations.
The matrices $G_i$ are maps between irreducible representations
of $Spin(4)$. A fuzzy $S^3$ will be defined by
matrices ${\hat G}_i$ whose matrix elements are those of  
$G_i$ in a direct sum of irreducible $Spin(4)$ representations 
related by $G_i$.  
The irreducible representations in this direct sum will be
fixed by requiring $\sum {\hat G}_i{\hat G}_i$ to be 
proportional to the identity.

Since $Spin(4) = SU(2) \times SU(2)$, the irreducible representations
are labelled by a pair of spins $(j_l, j_r)$.  The fundamental 
spinor representation $V$ of $Spin(5)$ decomposes under $Spin(4)$ into
$(1/2,0) \oplus (0,1/2) = P^+V \oplus P^-V$,  where $P^+$ and $P^-$
are positive and negative chirality projectors.   
The symmetric tensor representations
of $Spin(5)$, $Sym^{n} (V)$,  decompose into  irreducible representations 
of $Spin(4)$ as follows;
\eqn\decomp{ Sym( V^{\otimes n } ) = \oplus_{k=0}^{n} V_{( { n-k\over 2} ,
 { k \over 2 })   } } 
This can be proved by observing 
 \eqn\pf{  Sym( V^{\otimes n } ) = P_n V^{\otimes n } 
 = P_n ( P_+ + P_- )^{\otimes n}  V^{\otimes n } 
 = P_n \sum_{k} ({P_+}^{\otimes k} {P_-}^{\otimes n-k})_{sym} 
V^{\otimes n} } 
The dimension of $Sym^{n} (V) $ is $ {( n+1 ) ( n+2 ) (n+ 3 ) \over
 6} $. It is easy to check that the above decomposition is consistent 
 with the dimensions of the representations. 
\eqn\smrps{ 
\sum_{k= 0 }^{n}   ( n-k + 1 ) ( k +1 ) = { ( n +1) ( n+2 ) ( n+3 )
 \over 6 }  } 

Since $\Gamma^i P_+ = P_-\Gamma^i$ and $\Gamma^i P_- = P_+ \Gamma^i$,
$G_i$ is a map between  different 
irreducible representations of $Spin(4)$ and
vanishes when restricted to a particular irreducible representation. 
The fuzzy $S^3$ will be defined the matrices 
\eqn\proj{  {\hat G}_i = P_{\cal R} G_i P_{\cal R}}
where $P_{\cal R}$ is the projector to a reducible representation
${\cal R}$.  
We require that $\sum_{i=1}^4 G_iG_i$ is proportional to the identity 
within ${\cal R}$. 
One can show that $\sum_{i =1}^4 G_iG_i$ commutes with the
$Spin(4)$ generators $\sum_l \rho_l(\sigma_{jk})$ (see the appendix), 
where $\sigma_{ij} = \Gamma_i \Gamma_j - \Gamma_j \Gamma_i $.   
By Schur's lemma,  $\sum_iG_iG_i$ is proportional to the identity 
within each irreducible representation\foot{For $n=1$,  $G_iG_i$ 
is equal to $1$ for every $i$,  without
any need to sum on $i$.  For large $n$ one must sum over $i$ to obtain
something which is proportional to the identity.
This is fortunate,  since otherwise one could not obtain a classical
$S^3$ in the large $n$ limit.}
However, the proportionality
factor may differ between them.  We must 
therefore take ${\cal R}$ to be the direct sum of irreducible 
representations related by $G_i$ and having the same $\sum_i G_i G_i$.

In the representation $({n+k \over 2}, {n-k\over 2})$, 
one finds (see the appendix) that
\eqn\rtwo{ \sum_i {\hat G}_i{\hat G}_i = 4(n + nk - k^2).}
This is symmetric under $k\rightarrow n-k$. However
there is no other degeneracy.  Thus  
$\sum G_iG_i$ is proportional to the identity
in the representation
$({k\over 2}, {n-k\over 2}) \oplus   ({n-k\over 2},{k\over 2})$. 
For ${\hat G}_i$ to be nontrivial in this representation,  
$G_i$ must be a map between it's irreducible components.  This gives 
$k = n-k \pm 1$ and
\eqn\rep{ {\cal R} = {\cal R}_+ \oplus {\cal R}_- =  
({n+1 \over 4},{n-1 \over 4}) \oplus 
({n-1 \over 4}, {n+1 \over 4})}
The dimension of ${\cal R}$ is 
\eqn\dims{N = {1\over 2}(n+1)(n+3).}   

The restriction to this particular sum of $Spin(4)$ irreducible 
representations essentially amounts to considering a subspace 
of the fuzzy $S^4$ of \cas, 
which previously included $Spin(4)$ representations with all values of $k$.
The fuzzy $S^4$ has 
\eqn\foursph{\sum_{\mu =1}^5 G_{\mu}G_{\mu} = n^2+4n.}
For the fuzzy $S^3$  in the representation ${\cal R}$, 
we find (see the appendix) that
\eqn\rad{\sum_{i=1}^4{\hat G}_i{\hat G}_i = n^2 + 4n - 1}
Note that in the large $n$ limit,  the leading term is the 
same for the $S^3$ and the $S^4$.
Thus the fuzzy $S^3$ is a ``great sphere'' on an equator of the fuzzy $S^4$.
One can easily check that $G_5 G_5 = 1$  
when restricted to the representation
${\cal R}$.  

A general form for a fuzzy $S^3$ which approaches a classical
$S^3$ in the large $n$ limit is given by matrices
\eqn\gen{\phi_i = {1\over n}(a{\hat G}_i + ib{\hat G}_i 
(P_{{\cal R}_+} - P_{{\cal R}_-})) = {1\over n} 
(a{\hat G}_i + ib{\hat G}_i {\hat G}_5) }
which is the natural generalization of the $n=1$ expression
$\phi_i = a\Gamma_i + ib\Gamma_i\Gamma_5$. 
Since
\eqn\antic{ {\hat G}_i{\hat G}_5 = 
        - {\hat G}_5 {\hat G}_i,}
it follows that 
\eqn\radf{ R^2 = \phi_i\phi_i = (a^2 + b^2){(n^2 + 4n - 1)\over n^2}.}
The sphere is classical in the large $n$ limit because
$\phi_i$ has eigenvalues of order 1,  whereas 
\eqn\commut{[\phi_i,\phi_j] = {( a^2 + b^2 )\over n^2} 
P_{\cal R} \sum_l \rho_l(\sigma_{ij}) P_{\cal R}}
has largest eigenvalue of order $1\over n$.

\newsec{Large $N$ equations.  } 
We now insert the ansatz 
$\phi_i ={1\over n}( a{\hat G}_i + ib{\hat G}_i{\hat G}_5)$
into the equations of motion.  For the tachyon,  we make the 
ansatz 
\eqn\tachy{T = c + d(P_{{\cal R}_+} - P_{{\cal R}_-}) 
= c + d { \hat G_5}   }
With this ansatz,  the relations \fnfp\ and \exprs\
still apply with $\gamma_5 \rightarrow P_{{\cal R}_+} - P_{{\cal R}_-} = 
{\hat G}^5$.
Note that for $c = b = 0$,  the tachyon behaves like the fifth 
embedding coordinate of 
a fuzzy $S^4$, restricted to a particular $Spin(4)$ representation. 
Considering the class of solutions with $c=0$,  and
using properties described in the appendix,  one 
finds the following equations of motion.

The $ T$ equation of motion is; 
\eqn\teqn{\eqalign{
%& \omega \bigl[ \mu + {  \mu (n+1)  ( n+5 ) \over ( n(n+4) - 1 )^2 } R^4 
%+ 2d^2 { ( n+1) ( n+3) \over { n(n+4) - 1 }} R^2  \right) \cr 
	& ~~ {\hat G}^5 \bigl[ \mu \lambda + 
        \mu \lambda { (n+1) (n+5) \over { (n(n+4) -1)^2 }} R^4
	+ 2d^2\lambda {(n+1)(n+3) \over { n(n+4)-1 } } R^2  \cr 
	& + 4dq { (n+1) (n+3) \over { n(n+4) -1 } } 
 	- 4h e^{-d^2/8} 
	{ (n+5) (n+1)\over ( n(n+4) -1)^2 }R^4 \bigr] = 0  \cr }}
%After setting $ \omega =0 $ 
% which can be achieved by setting 
% $c=0$ and taking advantage of the 
% fact that $ f^{\prime} (T) - f^{\prime} (-T) = 0 $, 
% we are left with : 
%\eqn\teqaf{\eqalign{ &  
%\mu \lambda + 
%{ 2d^2\lambda (n+1)(n+3) \over { n(n+4)-1 } } R^2 
%+ \mu \lambda { (n+1) (n+5) \over { (n(n+4) -1)^2 }} R^4 \cr 
%&  4dq { (n+1) (n+3) \over { n(n+4) -1 } } R^2 - 4h { (n+5) (n+1) \over
%( n(n+4) -1)^2 } R^4  =0  \cr } }
The $ \phi$ equations of motion are; 
\eqn\phieq{ 
\left( d^2q + { 3 \mu q \over 2 } { R^2 \over { n(n+4 ) -1 }}
- { 2hd e^{-d^2/8}} {  R^2 ( n+2) \over { n(n+4) -1 }} \right) \phi^{i} = 0 }
 
\newsec{ Solutions }

We discuss here some properties 
of solutions to the above equations.  
%We will first 
%consider $n=1$. Then we will look at large 
While these solutions certainly 
have some of the requisite physical properties, 
we have no convincing identification with integral 
numbers of 
spherical D3-branes, as we will elaborate in the following.

At large $n$ the equations take the form : 
\eqn\sln{\eqalign{ & \mu \lambda + 2 d^2 \lambda R^2 + { \mu \lambda
 \over n^2} R^4 - { 4 h e^{-d^2\over 8} R^4 \over n^2 } 
+ 4dq R^2 = 0 \cr 
& d^2 q + { 3 \mu q \over 2 }{ R^2 \over n^2 } 
- 2 h e^{-d^2/8} d { R^2 \over n} =0.
 \cr }} 
We consider solutions with $ d = { \tilde d \over n }$ 
and $ R^2 = { \tilde  R}^2 n $  
with $ \tilde d $ and $ \tilde R$ finite in the large $n$ limit. 
 For the potential $  V =  e^{ -{ d^2 \over 4 } } $, one has
$\lambda = - {\tilde d}/2n$, and  $ \mu q = 1 $ in the 
 large $ n $ limit.
The leading terms in the equations of motion at large $n$ give
\eqn\lnsoln{ \tilde d = { 3 \over 4 h }}
 and 
\eqn\solnt{ \hat R^2 =  { {\tilde d} \over \mu h} }
Since $ \mu $ is positive, 
this gives positive $R^2$. 
 
For finite $n$, the  energy is given by 
\eqn\energn{\eqalign{  
E =  & q { ( n+1) (n+3) \over 2 } \bigl[ \mu + { ( n+1) (n+5) \over
( n(n+4) -1 )^2 } \mu R^4 + 2 { ( n+1) ( n+3) \over n(n+4) - 1 }  d^2
R^2 \bigr ] \cr 
& - 2dh { (n+1)^2 (n+3) (n+5) \over n^4 } \cr }}  
Similar remarks as in the $ n=1$ case 
regarding stability and corrections apply here.

\subsec{ Ramond-Ramond couplings} 

The relation to spherical D3-branes 
 can be  explored by a simple argument 
 analogous to the discussion of 
 fuzzy four-spheres \cas.  The net brane-charge is zero,
however one can determine the number of spherical D3-branes
by computing a charge contribution from a single hemisphere.
  
Recalling \kluson\  that the 3-brane potential 
couples to the zero-brane action through the term 
\eqn\cpl{ 
\int dt [\phi_1,\phi_2] [\phi_3,T] C_{0123}  
}  
The calculation of the charge of a single hemisphere is closely 
 analogous to that which gives 
 the 4-brane charge in the case
 of zero-branes polarizing to fuzzy 4-spheres
 \cas. 
In our case we have 
\eqn\charge{ {  ( 2 \pi )^4   \over 8 \pi^2 } \epsilon_{ijkl} 
	Tr_{1/2}\phi^{j} \phi^{k} [ \phi^{l}, T ] } 
where $Tr_{1/2}$ indicates that the trace is evaluated 
 in the subspace for which the eigenvalues of 
 $ \phi^{i} $ are positive. At large $n$,  one finds this to be equal to  
 $ 2d $ times the volume of the $\phi^i >0$ 
 hemisphere,  
 where $d$ is the quantity which enters our ansatz for the tachyon \tachy.  
\eqn\hchr{\eqalign{& 
2\pi^2 \epsilon_{ijkl}Tr_{1/2} \phi^{j} \phi^{k} [ \phi^{l}, T ]  
= \cr & 4 \pi^2 d {(n+2) \over n^3} Tr_{1/2}G^i  R^3\cr  
& \rightarrow d 4\pi^2 R^3 = 2d Vol_{1/2}.}}
Thus it appears that $d$ is proportional to the number of D3-branes.

Alternatively,  one may compute the dipole coupling of
of a spherical D3-brane to the Ramond-Ramond four-form and compare it
to the coupling of the D0 solution.
The couping of $Q_3$ spherical D3-branes to the a constant RR $5$-form 
field strength of the form  \ff\ is given by 
 \eqn\coup{ \eqalign{
 & \mu_3 Q_3 \int dt \int_{S^3}  PB( C^{(4)} ) = \cr 
 & \mu_3 Q_3 \int dt \int_{S^3 } h \epsilon_{ijkl} X^{i}
  {\partial X^j \over \partial \sigma^{\alpha}}
{\partial X^k \over \partial \sigma^{\beta}}
{\partial X^l \over \partial \sigma^{\gamma}}
 d{\sigma^{\alpha}} \wedge
 d{\sigma^{\beta}}\wedge d{\sigma^{\gamma}} = \cr
& 2\pi^2 \mu_3 Q_3 h R^4 }}  
On the other hand the coupling arising for the D0 configuration
is proportional to  
\eqn\dzerocoup{h
\mu_0 \int dt  \epsilon_{ijkl} Tr  
        \left[  [\phi^i,T] \phi^j \phi^k \phi^l \right]}  
which for our ansatz \gen \tachy becomes
\eqn\oura{\eqalign{& \int dt  \epsilon_{ijkl} Tr  
        \left[ 2d \phi^i \phi^j \phi^k \phi^l \right] \cr
	& =  2 d h R^4 {(n+1)^2(n+3)(n+5) \over (n^2 + 4n -1)^2
 }}}
or $2dhR^4$ for $n\rightarrow\infty$.

 Therefore (up to a numerical factor) the quantity $d$ which 
 appears in the 
 Tachyon ansatz \tachy, would
 be identified with the 
 number of  spherical three-branes
 if such a dual description existed. 
 However, for the large $ n$ solutions which we have found, 
 this quantity is zero as $ n \rightarrow \infty $. Thus 
 we cannot identify 
 these large $n$ solutions with spherical D3-branes. 
 Note that for the fuzzy $S^3$ ansatz, there are continuous 
 classes of solutions,  with
 $d$ depending on $h$.  In this case,  one would
 never expect a quantization of the dipole moment consistent
with a description in terms of integer numbers of D3-branes.
 Perhaps this situation changes if one considers probe D0-branes
in a consistent IIB supergravity background,  such as 
$ADS_5 \times S_5$ with quantized five-form flux on the $S^5$.
The polarization of stable branes in an a consistent AdS background
was considered in \mst.

%if what is said below were true,  the just taking the large n
%limit would give D3 branes on classical three-spheres.  So it
%is seems wrong
%At finite $n$,  the dipole moment does not vanish,  however
%for finite $n$ it is does not makes sense to compare
%with D3-branes on a classical sphere.
%Perhaps at finite $n$,  there is a dual description in 
%terms of D3-branes on a non-commutative $S^3$,  with sphalerons
%playing the role of the unstable D0-branes. 
%The relation between sphalerons and D0-branes has been pointed out   

\newsec{ Other odd non-commutative spheres } 

 The basic ingredients that went into our 
 construction of the non-commutative 
 three-sphere  admit a number of generalizations. 
 To understand this it is useful 
 to reformulate our discussion of the 
 non-commutative 3-sphere in a more 
 abstract form. 
 Essentially we needed objects $X^{\mu}$
 as matrices in some family of 
 representations 
 of $SO(4)$ parametrized by an integer $n$. They were 
 identified with  maps between  representations
 $R_1$ and $R_2$. $R_1$ was an irreducible 
 representation of $SO(4)$ of positive 
 chirality, i.e a representation 
 where $ \Gamma_5$ takes eigenvalue 
$1$.  $R_2$ was a representation 
 of negative chirality, i.e a representation 
 where $ \Gamma_5$ takes eigenvalue $-1$. 
We needed a non-zero coupling between 
$R_1$, $R_2 $ with the vector representation 
of $Spin(4)$. We may think of $X$ as the Clebsch-Gordan 
coupling $ R_1 \otimes R_2$ to the vector representation. 

Two natural generalizations are suggested 
 by the above. One is to consider odd 
 spheres of other dimensions. 
For example for $5$-spheres we 
 would like to take two representations 
 of $Spin(6)$ satisfying the above conditions. 
 Since $ Spin(6) $ is isomorphic to
$SU(4)$, we can exploit this to give a simple 
 description of the desired representations. 
The two chiral four dimensional  representations
  of $Spin(6)$ can be identified with 
  the fundamental and the anti-fundamental of 
 $SU(4)$.  These are represented in Young Tableau 
 notation by a single box, and a column of three 
 boxes respectively. We can take the representation
 $R_1$ to be given by the Young tableau with 
 $n$ columns of length $3$, and $n-1$ columns 
 of length $1$. The representation $R_2$ can be taken to
 be given by the Young Tableau with $n-1$ columns 
 of length $3$ and $n$ columns of length $1$. 

 Another generalization would to be consider 
 q-deformation. 
 We could take as our definition 
 of $X$ the Clebsch-Gordans of 
 the appropriate quantum group. 
  It would be interesting to compare 
  such a construction with the quantum spheres
  of \frt\  for example.

\newsec{ Summary and Outlook } 

We have found matrices defining non-commutative 
 3-spheres with a $Spin(4)$ invariance.
This construction is applicable for
 matrices of size
 $N = { ( n+ 1 ) ( n+3 ) \over 2 }$, 
 where $ n$ is odd. 
These three-spheres can be viewed as
a subspace of the non-commutative four-sphere
defined in \cas.
We have also proposed a general construction 
of higher dimensional odd fuzzy spheres along 
these lines.  While the construction of \cas\
gave a description of non-commutative even
spheres as solutions to Matrix Theory
(see also \ho\ for a variety of fuzzy spheres
appearing in Matrix Theory), the description of 
fluctations of fuzzy spherical branes is left
as an open problem.  This should be related
to the problem of giving a complete 
characterization of the algebra of functions
and differential calculus on the 
non-commutative sphere.  Likewise, we leave for 
future research the description, in terms of 
non-commutative geometry, of the fluctuations 
of the solutions we described.
The techniques of \cola\ may be useful here.
 
We have obtained unstable solutions of the IIB D0-brane 
equations of motion in the presence of a background 
RR-5 form background.  The adjoint scalars are 
given by the matrices defining the fuzzy $S^3$, and the 
tachyon is related to
the fifth matrix coordinate involved in the definition of
the fuzzy $S^4$ in which the $S^3$ is embedded.
The tachyon as an extra dimension has 
been hinted at in several papers \horav\ 
and the full significance of this intriguing 
geometrical behaviour remains to be understood.

At finite $n$,  the solutions we find carry an apparent 
D3 dipole moment.
In the large $n$ limit, the non-commutative 
$S^3$ becomes a classical $S^3$ and  the dipole moment vanishes.
Thus, contrary to what one may have expected, 
we can not view these solutions
as spherical D3-branes.  However, this may simply be a 
consequence of having considered an inconsistent
supergravity background. 
Note that if one did find solutions
which could be regarded as D3-branes on a classical $S^3$, 
then it seems likely that the D0-branes could be described by
sphalerons in  a dual $D3$ description.  The correspondance
between sphalerons and unstable branes has been discussed,
in a different context in \refs{\hhsphal, \dgsphal}.  

It would be very interesting to understand the
mechanism which might lead 
to D3-charge quantized in the standard way. The
choice of a consistent supergravity background  
mentioned above, the geometrical nature of the
tachyon in relation to fuzzy 4-spheres, 
as well as detailed properties of the tachyon
dependent couplings to 
Ramond-Ramond forms in unstable branes might 
be expected to
enter the story.

Odd spheres also appear in the ADS/CFT context 
  and non-commutative versions based on quantum groups 
  have been studied in the context of the stringy exclusion
  principle \malstrom\antram\hrt. The investigation 
 of the relation between quantum 3-sphere 
 constructions and the non-commutative 3-sphere in this paper 
 is  left for the future. Techniques like those 
 of \hidclass\ where a connection between fuzzy 2-sphere 
 and q-2sphere was established may be useful.  
 A different context where 
 odd non-commutative spheres have been discussed recently 
 is \hamo. A relation between non-commutative 
 $S^4$ and quantum $S^3$ also appears in \cola.
 Earlier discussions of fuzzy spheres appear 
 in \fro. It will be undoubtedly illuminating to undesrtand the 
 physical and mathematical relations 
 between these different constructions of non-comuutative 
 spheres.

%%%%%%%%%%%%%%%%%%%%%%%%%%%%%%%%%%%%%%%%%%%%%%%%%%%%%%

\bigskip

\noindent{\bf Acknowledgements:}
 We wish to thank for pleasant discussions Steve Corley, 
 Amihay Hanany, Antal Jevicki,  
  David Lowe, Robert Myers, Jan Troost and  Barton Zwiebach. 
 This research was supported  in part by DOE grant  
 DE-FG02/19ER40688-(Task A). 

\newsec{ Appendix }

We describe here some useful formulae.
 In this section  Roman indicies are taken
to run from $1$ to $4$ and Greek indicies to run from $1$ to $5$. 
Consider the symmetric tensor product representations
of $Spin(5)$ : 
\eqn\symt{ Sym(V^{\otimes n}) = (V_1 \otimes V_2 \otimes V_3 \cdots 
        \otimes V_n)_{sym} }   
where each $V_{\alpha}$ is a $4$ dimensional spinor and 
the symmetrization is as defined in \defp . 
The generators of $Spin(5)$ are 
\eqn\genrt{ G_{\mu\nu} = \sum_{l = 1}^n \rho_l(\sigma_{\mu\nu}) =
        \sigma_{\mu\nu}\otimes 1\otimes 1 \cdots +
        1\otimes \sigma_{\mu\nu} \otimes 1 \cdots}
It is easy to verify that
\eqn\comt{ \left[ \sum_{\mu =1}^5 \Gamma^{\mu} \otimes \Gamma^{\mu} \otimes 1
\otimes 1 \cdots, G_{\mu\nu} \right] = 0}
Thus in any irreducible representation of $Spin(5)$ the matrix    
$\sum_{\mu =1}^5 \Gamma^{\mu} \otimes \Gamma^{\mu} \otimes 1
\otimes 1 \cdots$ is proportional to the identity.  
In the symmetric tensor representations,  this proportionality 
factor is one,  as can be verified by
multiplying the whole expression on the left by
$\sum_{\mu =1}^5 \Gamma^{\mu} \otimes \Gamma^{\mu} \otimes 1
\otimes 1 \cdots$ twice and deriving a cubic equation for the
constant using $\Gamma$-matrix identities. 
Using this,  one finds that 
matrices $G^{\mu}$ defined in the symmetric tensor 
 representations of $Spin(5)$ by

\eqn\matr{G^{\mu} = \sum_{l=1}^n \rho_l(\Gamma^{\mu} ) = 
(\Gamma^{\mu} \otimes 1 \otimes 1  \cdots  ~
+ 1\otimes \Gamma^{\mu} \otimes 1 \cdots  ~~ + \cdots ~)_{sym}}
satisfy the property 
\eqn\smmu{ \sum_{\mu =1}^5 G^{\mu}G^{\mu} = n^2+4n .}

The symmetric tensor representations of $Spin(5)$ decompose under
$Spin(4)$ as in \decomp\pf;
\eqn\dcmp{Sym( V^{\otimes n } ) 
         = P_n \sum_{k} ({P_+}^{\otimes k} {P_-}^{\otimes n-k})_{sym} 
        V^{\otimes n} .}
 The matrices of the form
\eqn\frm{ \sum_{i=1}^4 \Gamma^i \otimes \Gamma^i \otimes 1 \otimes 1
\cdots = 1 - ~~\Gamma^5 \otimes \Gamma^5 \otimes 1\otimes 1 \cdots}
commute with all the $Spin(4)$ generators $G_{ij}$.
If one restricts to one of the $Spin(4)$ irreducible representations
labelled by $k$,  then this quantity is  proportional
 to the identity by Schur's Lemma.  
Using \frm\ it can be shown that
\eqn\radthr{\sum_{i=1}^4 { G}^i{ G}^i = 4(n^2 + nk - k^2), }
when restricted to the representation labelled by $k$. 

Let us now consider the reducible $Spin(4)$ representation
$({n+1\over 4}, {n-1\over 4}) \oplus ({n-1\over 4}, {n-1\over 4})$
which is the sum of representations with $k=(n+1)/2$ and $k = (n-1)/2$
with  odd $n$.  We write the restriction of $G^{\mu}$ to this reducible
representation as ${\hat G}^{\mu}$. 
Taking $T= c + d{\hat G^5}$ and $f(T) = q + r {\hat G}^5$,  one 
obtains the following relations for the terms which appear in 
the equations of motion.

\eqn\lst{\eqalign{ & {\hat G}^i{\hat G}^i = n^2+4n -1 \cr
   & [{\hat G}^i,{\hat G}^j][{\hat G}^i,{\hat G}^j] =  
     -4(n+1)(n+5) \cr
   &[{\hat G}^i, T][{\hat G}^i, T] = -2 
    d^2(n+1)(n+3) \cr
   & \left[ {\hat G}^i, \{[{\hat G}^i, T],
    f(T)\}\right] = 4dq(n+1)(n+3) {\hat G}_5  \cr
   & \epsilon_{ijkl}{\hat G}^i{\hat G}^j{\hat G}^k{\hat
     G}^l = 2(n+1)(n+5)   {\hat G}_5 \cr 
   & \left[  \{ f(T),[{\hat G}^i,T]\}, T\right] =
        8d^2q{\hat G}^i \cr
   & \left[ \{f(T),[{\hat G}^i,{\hat G}^k]\},{\hat
     G}^k\right]=
       24q{\hat G}^i \cr
   & \epsilon_{ijkl}\{  [{\hat G}^j, {\hat G}^k], [{\hat
     G}^l,T]\}
        = 16 d (n+2){\hat G}^i
  }}

\listrefs
\end